\documentstyle[12pt]{article}

\input epsf.sty
\newcommand{\insertfig}[2]{\mbox{\epsfxsize=#1cm \epsfbox{#2.eps}}}
\newcommand{\Sym}{\mathop{\mbox{\large\bf S}}}
\def\1{\hbox{{1}\kern-.25em\hbox{l}}}

\begin{document}

\begin{titlepage}

\begin{flushright}
DOE/ER/40762-252 \\ [-2mm]
UMD-PP\#02-043 \\
\end{flushright}

\centerline{\large \bf Chiral structure of nucleon gravitational form factors}

\vspace{15mm}

\centerline{\bf A.V. Belitsky, X. Ji}

\vspace{5mm}

\centerline{\it Department of Physics}
\centerline{\it University of Maryland at College Park}
\centerline{\it College Park, MD 20742-4111, USA}

\vspace{20mm}

\centerline{\bf Abstract}

\vspace{1cm}

We study the low momentum behavior of nucleon gravitational form factors in the
framework of the heavy baryon chiral perturbation theory. At zero recoil they
determine the momentum and spin apportion between nucleon constituents. Our
result provides an insight into the response of the nucleon's pion cloud to an
external weak gravitational field and establishes a theoretical framework for
extrapolation of experimental and lattice data on the nucleon form factors to
zero momentum transfer. We also discuss form factors corresponding to higher-rank
tensor currents related to the moments of generalized parton distributions.

\vspace{60mm}


\vspace{5mm}


\end{titlepage}

{\bf 1.} The fundamental problem of hadronic physics is the understanding
of the structure of the nucleon, --- the building block of our planet
and its inhabitants, --- in terms of quark and gluonic degrees of freedom.
The nucleon's mass and momentum content is one of the issues being under
study for decades. Higgs mechanism giving quarks their masses generates only
a tiny fraction of the total nucleon mass, the bulk of which comes entirely
from the complicated dynamics of colored glue. This idea was supported by
deeply inelastic scattering experiments which revealed that quarks and
gluons are equally responsible for the energy-momentum structure of the
nucleon with both of them carrying about $50\%$ of its momentum. A similar
study has to be addressed for the spin content of hadrons. The nucleon
being a composite particle builds its spin from the angular
momenta\footnote{For a gauge particle the decomposition of its angular
momentum into its spin and orbital components is not possible. The
hand-waiving argument goes as follows: In quantum mechanics, the wave
function of a spin-$s$ particle is a symmetric rank-$2s$ spinor having
$(2s + 1)$ components which transform into each other under the rotation
of the coordinate system. The orbital wave function is related to the
coordinate dependence of wave functions and is given by the spherical
harmonics of order $l$ for the angular momentum $l$ of the system.
Therefore, in order to distinguish clearly between the spin and orbital
momentum, the spin and coordinate properties must be independent. As it
is obvious, this condition is not fulfilled for gauge particles whose
description in terms of field operators inevitably involves a gauge
condition. For instance, in the Coulomb gauge the gluon wave function
is given by the three-potential $\mbox{\boldmath$A$}^a (x)$ equivalent
to the second rank spinor which, however, is a subject for the gauge
condition ${\rm div} \mbox{\boldmath$A$}^a (x) = 0$. As a result the
coordinate dependence of the vector cannot be independently defined for
each of its components and leads to the inability to separate the spin
and orbital degrees of freedom.} of its constituents $J_{q,g}$, i.e.,
in the case of quarks from their spin $\Delta \Sigma / 2$ and orbital
motion $L_q$,
\begin{equation}
\label{SpinSumRule}
\frac{1}{2} = \frac{\Delta \Sigma}{2} + L_q + J_g \, .
\end{equation}
A next natural question arises about the numerical apportion between these
contributions. The naive quark model (QM) attributes the whole of the nucleon
spin to the sum of the ones of quarks $\Delta \Sigma_{\rm QM} = 1$
\cite{Clo79}. However, the experimental data tells the opposite: only a small
fraction of the total spin is carried by quarks $\Delta \Sigma_{\rm exp} =
0.16 \pm 0.08$ \cite{FilJi01}. Thus in order to test the spin sum rule
(\ref{SpinSumRule}) one has to have access to the other sources of the
nucleon spin $L_q$ and $J_g$. The nucleon's angular momentum is expressed
in QCD by a hadronic matrix element of the coordinate-space moment of the
quark and gluon momentum densities given by specific components of
the energy-momentum tensor. At this point recall that the magnetic
moment of a particle is the space moment of the electric current flow and
can be measured as a matrix element of the electric current at non-zero
momentum transfer, i.e., in elastic electron-nucleon scattering
by accessing electromagnetic form factors. The same methodology is
applicable in the current circumstances so that the missing orbital
momentum parts can be deduced from off-forward matrix elements of the
QCD energy-momentum tensor. Thus, in order to understand the nucleon
spin content one has to study nucleon gravitational form factors.
These functions, which arise in the nucleon scattering off a weak
gravitational field, are measurable experimentally in exclusive processes
such as deeply virtual Compton scattering \cite{Ji98,Rad96,BelMulKir01}
and hard exclusive meson production \cite{ColFraStr96,GoePolVan01}.
Moreover, they can be simulated on a lattice. As we just said, apart
from the interest in their own right, gravitational form factors at zero
recoil determine the quark and gluon contributions to the angular momentum.
In practical determination of the latter either in laboratory or
lattice experiments one needs in all cases to perform an extrapolation
to the limit of zero momentum transfer $\Delta^2 = 0$ to get the
net effect. The experimental kinematics is restricted to low $t$-channel
momenta of order $0.1\ {\rm GeV}^2$ \cite{BelMulKir01} and, thus, favors
the applicability of the chiral perturbation theory ($\chi$PT) \cite{Wei96}
to unravel the small-momentum transfer behavior of the form factors in
question. On a lattice, the minimal achieved momentum is limited by the
size of the lattice. The first Monte Carlo simulations
\cite{MatDonLiuManMul00,GadJiJun01} were done with lattices
$16^2 \times 32$ where $\Delta^2_{\rm min} \sim 0.4 \ {\rm GeV}^2$. Thus
in order to enter the regime of validity of $\chi$PT one needs repeat
the simulations with bigger lattices. In the present Letter we compute
one-loop contributions to off-forward matrix elements of local quark and
gluon composite operators and thus determine their momentum transfer
dependence.

{\bf 2.} The spin and momentum structure of the nucleon is carried by
off-forward matrix elements of quark and gluon composite operators. In our
consequent presentation we will be concerned with the parity-even sector
which can shed some light on the spin structure of the nucleon. Under
chiral $SU (2)_L \otimes SU (2)_R$ group the operators transform either
as isovector $(3,1) \oplus (1, 3)$ or isoscalar $(1, 1)$ representations
and correspond to flavor non-singlet ${\cal R}^{q, a}$ and singlet
${\cal R}^{q, 0}$ quark operators, respectively,
\begin{equation}
\label{QuarkOperator}
{\cal R}^{q,A}_{\mu_1 \mu_2 \dots \mu_j}
= \Sym_{\mu_1 \mu_2 \dots \mu_j}
\bar\psi
\tau^A
\gamma_{\mu_1}
i\!\stackrel{\leftrightarrow}{\cal D}_{\mu_1}
i\!\stackrel{\leftrightarrow}{\cal D}_{\mu_2}
\dots
i\!\stackrel{\leftrightarrow}{\cal D}_{\mu_j}
\psi \, ,
\end{equation}
where $\tau^A = (\1 , \tau^a)$ with $\1 \equiv 1_{[2] \times [2]}$.
The operator of symmetrization and trace subtraction is defined as
follows, e.g., $\Sym_{\mu_1 \mu_2} t_{\mu_1 \mu_2} = \frac{1}{2!}
(t_{\mu_1 \mu_2} + t_{\mu_2 \mu_1} - \frac{1}{2} g_{\mu_1 \mu_2}
t_{\mu\mu})$. The isoscalar combination ${\cal R}^{q, 0}$ mixes
under the renormalization group evolution with gluonic operators,
\begin{equation}
\label{GluonOperator}
{\cal R}^{g}_{\mu_1 \mu_2 \dots \mu_j}
= \Sym_{\mu_1 \mu_2 \dots \mu_j}
G_{\nu\mu_1}
i\!\stackrel{\leftrightarrow}{\cal D}_{\mu_1}
i\!\stackrel{\leftrightarrow}{\cal D}_{\mu_2}
\dots
i\!\stackrel{\leftrightarrow}{\cal D}_{\mu_{j - 1}}
G_{\mu_j \nu} \, .
\end{equation}
The nucleon matrix element of these operators are parametrized via `form
factors' as follows
\begin{eqnarray}
\label{Decomposition}
\langle p_2 | {\cal R}^A_{\mu_1 \dots \mu_j} | p_1 \rangle
\!\!\!&=&\!\!\! \Sym_{\mu_1 \dots \mu_j}
\bar u (p_2) \tau^A \gamma_{\mu_1} u (p_1)
\left\{
p_{\mu_2} \dots p_{\mu_j} A_{j,j} (\Delta^2)
+
\dots
+
\Delta_{\mu_2} \dots \Delta_{\mu_j} A_{j,1} (\Delta^2)
\right\}
\nonumber\\
&+&\!\!\!
\Sym_{\mu_1 \dots \mu_j}
\bar u (p_2) \tau^A \frac{i \sigma_{\mu_1 \nu} \Delta_\nu}{2 M} u (p_1)
\left\{
p_{\mu_2} \dots p_{\mu_j} B_{j,j} (\Delta^2)
+
\dots
+
\Delta_{\mu_2} \dots \Delta_{\mu_j} B_{j,1} (\Delta^2)
\right\}
\nonumber\\
&+&\!\!\!
\Sym_{\mu_1 \dots \mu_j}
\frac{\bar u (p_2) \tau^A u (p_1)}{2 M}
\Delta_{\mu_1} \dots \Delta_{\mu_j} C_{j} (\Delta^2)
\, ,
\end{eqnarray}
where we have introduced the vectors $p \equiv p_1 + p_2$ and $\Delta = p_2
- p_1$ and assumed the normalization for hadronic states $\langle p | p
\rangle = 2 p_0 V$ and bispinors $\bar u u = 2 M$. We have reduced in the
above equation all tensor structures of the scalar Dirac bilinear containing
at least one vector $p_\mu$ by means of the Gordon identity to the vector
and tensor Dirac bilinears of the first two lines. There is an important
constraints on the matrix elements (\ref{Decomposition}) imposed by the
time-reversal invariance, namely, vanishing of Lorentz structures with odd
powers of the momentum transfer $\Delta$ \cite{Ji98}. We will keep this
condition in mind without implementing it explicitly in our subsequent
considerations.

The reduced matrix elements $A$, $B$ and $C$ are related to the moments
of parity-even quark or gluon generalized parton distributions
\cite{MulRobGeyDitHor94,Ji96,Rad96} measurable in deeply virtual Compton
scattering via
\begin{eqnarray}
\int_{-1}^1 d x \, x^{j - 1} H (x, \eta, \Delta^2)
\!\!\!&=&\!\!\! \sum_{k = 0}^{j - 1} \eta^k
A_{j, j - k} (\Delta^2) + \eta^j C_{j} (\Delta^2)
\, , \nonumber\\
\int_{-1}^1 d x \, x^{j - 1} E (x, \eta, \Delta^2)
\!\!\!&=&\!\!\! \sum_{k = 0}^{j - 1} \eta^k
B_{j, j - k} (\Delta^2) - \eta^j C_{j} (\Delta^2)
\, .
\end{eqnarray}
The angular momentum sum rule (\ref{SpinSumRule}) involves the form factors at
zero recoil entering their second moment \cite{Ji96}
\begin{equation}
\label{SpinRule}
J
=
\frac{1}{2}
\lim_{\Delta^2 \to 0}
\left\{
A_{2,2} (\Delta^2) + B_{2,2} (\Delta^2)
\right\} \, ,
\end{equation}
which are attributed to quarks or gluons depending on the contributing
composite operators.

We consider single-nucleon systems and, in order to have a consistent power
counting, use the formalism of the heavy baryon $\chi$PT which treats the
nucleon as a non-relativistic infinitely heavy particle \cite{JenMan91}. For
the purpose of a unique translation of the Lorentz-covariant matrix elements
in Eq.\ (\ref{Decomposition}) into non-relativistic ones its is convenient to
use the Breit reference frame \cite{BerKaiKamMei92,BerFeaHemMei98}. It is
defined by the condition $\mbox{\boldmath$p$} = 0$ which leads in turn to
$\Delta_0 = 0$, $p_0 = 2 M \sqrt{1 - \Delta^2/(4 M^2)} \equiv 2 M \delta$, and
$\mbox{\boldmath$p$}_2 = - \mbox{\boldmath$p$}_1 = \mbox{\boldmath$\Delta$}/2$.
In the Breit frame, the Dirac bilinears arising in Eq.\ (\ref{Decomposition})
are reduced via equations
\begin{eqnarray*}
&&\bar u (p_2) \gamma_\mu u (p_1)
= v_\mu \bar u_v (p_2) u_v (p_1)
+ \frac{1}{M} \bar u_v (p_2) [S_\mu, S \cdot \Delta] u_v (p_1)
\, , \\
&&\bar u (p_2) \frac{i \sigma_{\mu\nu} \Delta_\nu}{2 M} u (p_1)
= v_\mu \frac{\Delta^2}{4 M^2} \bar u_v (p_2) u_v (p_1)
+ \frac{1}{M} \bar u_v (p_2) [S_\mu, S \cdot \Delta] u_v (p_1)
\, ,
\end{eqnarray*}
to the ones constructed from the projected large components of the heavy baryon
bispinor $u_v (p_i) \equiv (1 + {\not\!v}) u (p_i) / \sqrt{2 + 2 (v \cdot p_i )/ M}$
normalized as $\bar u_v u_v = 2 M$. The velocity $v$ arises in the heavy-mass
decomposition of the incoming and outgoing nucleon momenta $p_{1,2} = M v + k \mp
\Delta/2$ and the residual momentum is defined as $k = \left( M (\delta - 1),
\mbox{\boldmath$0$} \right)$ when $v = (1, \mbox{\boldmath$0$})$. The spin
vector $S_\mu \equiv \frac{i}{2} \sigma_{\mu\nu} \gamma_5 v_\nu$ reduces
to the spin three-vector $S = (0, \mbox{\boldmath$\mit\Sigma$} /2 )$ in
the nucleon rest frame. Exploiting these results, the decomposition
(\ref{Decomposition}) transforms in the Breit frame to the equation
\begin{eqnarray}
&&\!\!\!\!\!\!\!\!\!\!
\langle p_2 | {\cal R}^A_{\mu_1 \dots \mu_j} | p_1 \rangle
= \! \Sym_{\mu_1 \dots \mu_j}\!\!
\bar u_v (p_2) \tau^A u_v (p_1) v_{\mu_1} \!
\left\{
(2 M \delta)^{j - 1}
E_{j, j} (\Delta^2)
v_{\mu_2} \dots v_{\mu_j} \!
+
\dots
+
E_{j, 1} (\Delta^2)
\Delta_{\mu_2} \dots \Delta_{\mu_j}
\right\}
\nonumber\\
&&\!\!\!\!\!
+\!
\Sym_{\mu_1 \dots \mu_j}
\frac{1}{M} \bar u_v (p_2) \tau^A [S_{\mu_1}, S \cdot \Delta] u_v (p_1)
\left\{
(2 M \delta)^{j - 1}
M_{j, j} (\Delta^2)
v_{\mu_2} \dots v_{\mu_j}
+
\dots
+
M_{j, 1} (\Delta^2)
\Delta_{\mu_2} \dots \Delta_{\mu_j}
\right\}
\nonumber\\
&&\!\!\!\!\!
+
\Sym_{\mu_1 \dots \mu_j}
\frac{\bar u_v (p_2) \tau^A u_v (p_1)}{2 M \delta^{-1}}
\Delta_{\mu_1} \dots \Delta_{\mu_j} C_{j} \, ,
\end{eqnarray}
where we have introduced a short-hand notation for the generalized electric-
and magnetic-like form factors
\begin{equation}
E_{j, k} \equiv A_{j, k} + \frac{\Delta^2}{4 M^2} B_{j, k}
\, , \qquad
M_{j, k} \equiv A_{j, k} + B_{j, k}
\, .
\end{equation}
Now we are in a position to address the calculation of the small-$\Delta$
dependence of these form factors. As it is obvious from the tensor
decomposition, one cannot compute form factors accompanying Lorentz tensors
involving several $\Delta_\mu$, i.e., $k < j - 2$, while restricting oneself
to next-to-leading order in the chiral expansion. Also note that $E_{1,1}$ and
$M_{1,1}$ correspond to the standard Sachs nucleon electromagnetic form factors
extensively discussed in the literature \cite{BerKaiKamMei92,BerFeaHemMei98}
and, therefore, their evaluation will not be repeated presently.

{\bf 3.} The calculational procedure is rather straightforward. First,
one constructs composite operators from the nucleon and pion fields
which match the quantum numbers of the ones on the quark-gluon level
(\ref{QuarkOperator},\ref{GluonOperator}) and adds them to the effective
Lagrangian \cite{ArnSav01,CheJi01}. Then one uses the heavy-baryon chiral
perturbation theory \cite{JenMan91}, see also the review \cite{BerKaiMei95},
for the computation of tree and one-loop contributions. For the case at hand,
one needs the leading order pion-nucleon Lagrangian
\begin{equation}
{\cal L} = \bar N_v
\left\{
i \, v \cdot {\cal D}
+ 2 g_A S \cdot {\cal A}
\right\}
N_v +
\frac{f_\pi^2}{4}
{\rm tr}
\left\{
\partial_\mu {\mit\Sigma} \ \partial_\mu {\mit\Sigma}^\dagger
\right\}
+ \lambda \,
{\rm tr}
\left\{ M_q
\left(
{\mit\Sigma} + {\mit\Sigma}^\dagger
\right)
\right\}
\, .
\end{equation}
The axial-vector and vector (in the covariant derivative ${\cal D}_\mu
\equiv \partial_\mu + {\cal V}_\mu$) pion potentials are
\begin{equation}
{\cal V}_\mu
\equiv
\frac{1}{2}
\left(
\xi^\dagger \partial_\mu \xi
+
\xi \partial_\mu \xi^\dagger
\right)
\, ,
\qquad
{\cal A}_\mu
\equiv
\frac{i}{2}
\left(
\xi^\dagger \partial_\mu \xi
-
\xi \partial_\mu \xi^\dagger
\right)
\, .
\end{equation}
Here ${\mit\Sigma}$
is constructed from the multiplet of the pion fields
\begin{equation}
{\mit\Sigma} \equiv \xi^2
= \exp \left( \frac{i}{f_\pi} \vec\pi \cdot \vec\tau \right) \, ,
\end{equation}
and $f_\pi = 93 \ {\rm MeV}$ the pion decay constant. Expanding to the first
non-trivial order in the pion fields we get for the axial and vector
potentials: ${\cal A}_\mu = - \frac{1}{2 f_\pi} \partial_\mu \, \vec\pi
\cdot \vec\tau + \dots$ and ${\cal V}_\mu = \frac{i}{(2 f_\pi)^2}
\varepsilon_{abc} \tau^c \pi^a \partial_\mu \pi^b + \dots$, respectively.
In the mass term, the coefficient $\lambda$ is related to the quark
condensate via $\lambda = - \frac{1}{2} \langle \bar \psi \psi \rangle$
and the quark mass matrix is $M_q = {\rm diag} (m_u, m_d)$.

Ultraviolet divergences generated by one-loop diagrams have to be absorbed
into coefficients of next-to-leading order counterterms in the chiral expansion.

Now we are in a position to discuss the construction of a basis of
hadronic twist-two operators corresponding to the quark and gluon ones
in Eqs.\ (\ref{QuarkOperator}) and (\ref{GluonOperator}). There are two
types of local hadronic operators: built either from the pions alone or
bilinear in the nucleon fields. The nucleon operators have the form
which mimics the tensor decomposition of the off-forward matrix elements.
The leading operators contributing to isoscalar and isovector combinations
of $E_{j,j}$ and $M_{j,j}$ are
\begin{eqnarray}
\label{BaryonOperators}
{\cal O}^{N, A}_{\mu_1 \mu_2 \dots \mu_j}
\!\!\!&=&\!\!\!
a^N_j (2 M)^{j - 1} \Sym_{\mu_1 \dots \mu_j}
v_{\mu_1} \dots v_{\mu_j}
\bar N_v \tau^A_{\xi +} N_v
\nonumber\\
&+&\!\!\!\ b^N_j (2 M)^{j - 1}
(- i \partial_\nu)
\Sym_{\mu_1 \dots \mu_j}
v_{\mu_1} \dots v_{\mu_{j - 1}}
\bar N_v \tau^A_{\xi +} \frac{[S_{\mu_j}, S_\nu]}{M} N_v
+ \dots
\, ,
\end{eqnarray}
where
\begin{equation}
\tau^A_{\xi +} \equiv \frac{1}{2}
\left(
\xi \tau^A \xi^\dagger + \xi^\dagger \tau^A \xi
\right) \, .
\end{equation}
The matching coefficients $a^N$ and $b^N$ from the partonic to the hadronic
level are unknown and have to be determined from experimental data. Subleading
operators are deduced from this expression by a mere replacement of $v$s in
the Lorentz structure by the derivatives. However, due to the time-reversal
symmetry restrictions one has to replace an even number of velocities. Here
and everywhere later, the nucleon operators are normalized as follows
$N_v (0) | p \rangle = u_v (p) | 0 \rangle$.

The pion sector has to be considered separately for the isoscalar and isovector
components. As we will see momentarily, one can essentially reduce the number
of operators at each level in the derivative expansion provided one uses all
underlying symmetries. A number of relations arises on the basis of the
following `magic' property of the Pauli matrices $\tau^2 \vec\tau \, \tau^2
= - \vec\tau\,{}^T$ which results into a useful conjugation property for the
nonlinear pion field ${\mit\Sigma}$,
\begin{equation}
\label{MagicRelation}
\tau^2 {\mit\Sigma}^\dagger \tau^2 = {\mit\Sigma}^T
\, .
\end{equation}

{\bf 4.} Let us start from the isoscalar sector which transforms as $(1,1)$
under chiral $SU(2)_L \otimes SU (2)_R$. The leading nucleon operators are
displayed in Eq.\ (\ref{BaryonOperators}) for $A = 0$ so that $\tau^0_{\xi + }
= \1$. Next, one can immediately conclude that the lowest local operator in
Eq.\ (\ref{QuarkOperator}) does not correspond to any pion operator as a
consequence of the unitarity of ${\mit\Sigma}$ and Eq.\ (\ref{MagicRelation}),
\begin{equation}
{\rm tr}
\left\{
{\mit\Sigma} \ \partial_\mu {\mit\Sigma}^\dagger
\right\} = 0 \, .
\end{equation}
Its generalization for an arbitrary odd $j = 2 k + 1$ reads
\begin{equation}
\Sym_{\mu_1 \mu_2 \dots \mu_j}
{\rm tr}
\left\{
\partial_{\mu_1} \dots \partial_{\mu_k} {\mit\Sigma} \
\partial_{\mu_{k + 1}} \dots \partial_{\mu_j} {\mit\Sigma}^\dagger\
\right\}
=
\frac{1}{2}
\Sym_{\mu_1 \mu_2 \dots \mu_j}
\partial_{\mu_1}
{\rm tr}
\left\{
\partial_{\mu_2} \dots \partial_{\mu_{k + 1}} {\mit\Sigma} \
\partial_{\mu_{k + 2}} \dots \partial_{\mu_j} {\mit\Sigma}^\dagger\
\right\}
\, .
\end{equation}
Therefore, all pion operators with an odd number of derivatives $j = 2 k + 1$
can be reduced to the ones containing an odd number of total derivatives, so
that we have
\begin{eqnarray*}
\Sym_{\mu_1 \dots \mu_j}
\left\{
\sum_{l = 0}^{k - 1} a^{\pi}_{j, 2 l + 1} \, ( -i \partial)^{2 l + 1} \,
{\rm tr}
\left\{
(- i \partial)^{k - l} {\mit\Sigma} \ (i \partial)^{k - l} {\mit\Sigma}^\dagger\
\right\}
\right\}_{\mu_1 \dots \mu_j} \, .
\end{eqnarray*}
Here, on the right-hand side of the equation, we imply that the Lorentz indices
are attached to the partial derivatives $\partial$ in a particular order, e.g.,
$\{\partial^3\}_{\mu_1 \mu_2 \mu_3} = \partial_{\mu_1} \partial_{\mu_2}
\partial_{\mu_3}$. Note, however, that matrix elements of these operators
violate the time reversal properties of the off-forward matrix elements in
question (\ref{Decomposition}) and thus cannot contribute.

Using the symmetry properties established above we can easily write composite
pion operators for even $j = 2 k$,
\begin{equation}
{\cal O}^\pi_{\mu_1 \mu_2 \dots \mu_j}
= f_\pi^2 \Sym_{\mu_1 \dots \mu_j}
\left\{
\sum_{l = 0}^{k - 1} a^{\pi}_{j, 2 l} \, ( - i \partial)^{2 l} \,
{\rm tr}
\left\{
( - i \partial)^{k - l} {\mit\Sigma} \ (i \partial)^{k - l} {\mit\Sigma}^\dagger\
\right\}
\right\}_{\mu_1 \dots \mu_j} \, .
\end{equation}
Since quantum numbers of hadronic operators match both quark and gluon ones in
Eqs.\ (\ref{QuarkOperator}) and (\ref{GluonOperator}), the difference between the
latter on the hadronic level arises only in the value of the matching coefficients
which appear in two species, $a^{\pi, q}_{j, k}$ and $a^{\pi, g}_{j, k}$. Below,
the index which attributes hadronic operators either to the quark or gluon sector
will not be displayed explicitly for brevity. Note that $a^\pi_{j, 0} \equiv a^\pi_j$
survive in the forward matrix elements and, thus, are measurable in conventional
deeply inelastic scattering experiments. For instance, $a_2^{\pi}$ is related to
the momentum fraction of the pion carried by quarks and gluons forming it. For an
on-shell pion at low normalization point $\mu^2 \approx 0.3 \ {\rm GeV}^2$ one has
\cite{GluReySch99} $a^{\pi,q}_2 = \langle x_q \rangle_\pi \approx 0.7$ and
respectively $a^{\pi,g}_2 = \langle x_g \rangle_\pi \approx 0.3$ due to the momentum
conservation. There is a number of pion operators with more than two ${\mit\Sigma}$
fields since their insertion does not alter the twist of the composite operator,
however, they do not contribute to the one-loop matrix elements we are interested
in and will be totally omitted.

To summarize, the isoscalar nucleon gravitational form factors, --- the second
moment of the generalized parton distributions $H$ and $E$, --- receive
contributions from the following pion and nucleon operators:
\begin{eqnarray}
{\cal O}^\pi_{\mu_1 \mu_2}
\!\!\!&=&\!\!\! f_\pi^2 a^{\pi}_2 \Sym_{\mu_1 \mu_2}
{\rm tr}
\left\{
\partial_{\mu_1} {\mit\Sigma} \ \partial_{\mu_2} {\mit\Sigma}^\dagger\
\right\}
\, ,
\\
{\cal O}^N_{\mu_1 \mu_2}
\!\!\!&=&\!\!\!
a^N_2 (2 M) \Sym_{\mu_1 \mu_2}
v_{\mu_1} v_{\mu_2}
\bar N_v N_v
+
b^N_2 (2 M)
(- i \partial_\nu)
\Sym_{\mu_1 \mu_2}
v_{\mu_1} \bar N_v \frac{[S_{\mu_2}, S_\nu]}{M} N_v
\nonumber\\
&+&\!\!\!
c^N_2 \frac{1}{2 M}
\Sym_{\mu_1 \mu_2}
(- i \partial_{\mu_1}) (- i \partial_{\mu_2})
\bar N_v N_v \, ,
\end{eqnarray}
and we have also included an operator with two total derivatives which
generates the structure $C_2 (\Delta^2)$.

%%%%%%%%%%%%%%%%%%%%%%%%%%%%%%%%%%%%%%%%%%%%%%%%%%%%%%%%%%%%%%%%%%%%%
%            Figure 1
%%%%%%%%%%%%%%%%%%%%%%%%%%%%%%%%%%%%%%%%%%%%%%%%%%%%%%%%%%%%%%%%%%%%%
\begin{figure}[t]
\begin{center}
\mbox{
\begin{picture}(0,55)(210,0)
\put(0,0){\insertfig{15}{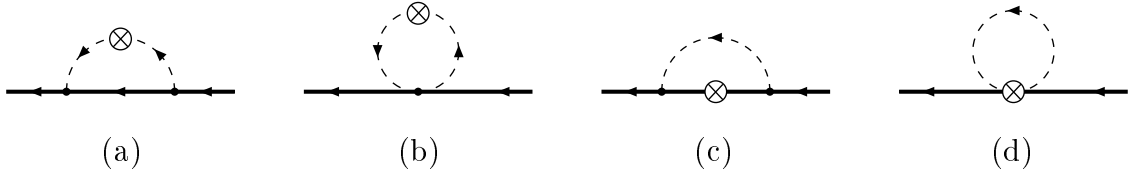}}
\end{picture}
}
\end{center}
\caption{\label{One-loop} One loop diagrams contributing to the matrix
elements of the twist-two operators. The self-energy insertions into
the external line, which are not displayed explicitly, have to be added.}
\end{figure}
%%%%%%%%%%%%%%%%%%%%%%%%%%%%%%%%%%%%%%%%%%%%%%%%%%%%%%%%%%%%%%%%%%%%%

A calculation of the diagrams (a) and (c), (b) and (d) do not contribute
for isoscalar operators, gives
\begin{eqnarray}
\label{IsovectorFFs}
E_{2, 2} (\Delta^2)
\!\!\!&=&\!\!\!
a^N_2 +
3 a^{\pi}_2 \frac{g_A^2}{64 \pi f_\pi^2 M}
\left\{
\left( 2 m^2 - \Delta^2 \right)
\int_0^1 dx \, \sqrt{m^2 (x)}
+
\frac{4}{3} m^3
\right\}
\, ,\nonumber\\
M_{2, 2} (\Delta^2)
\!\!\!&=&\!\!\!
b^N_2
+
3 \frac{g_A^2}{(4 \pi f_\pi)^2}
\int_0^1 dx \, \left( a^{\pi}_2 m^2 (x) - b^N_2 m^2 \delta (x) \right)
\left\{ \ln \frac{m^2 (x)}{\Lambda^2_\chi} - 1 \right\}
\, , \nonumber\\
C_{2} (\Delta^2)
\!\!\!&=&\!\!\!
c^N_2 +
3 a^{\pi}_2 \frac{g_A^2}{16 \pi f_\pi^2}
M \left( 2 m^2 - \Delta^2 \right)
\int_0^1 dx \, \frac{x (1 - x)}{\sqrt{m^2 (x)}}
\, ,
\end{eqnarray}
where $m^2 (x) \equiv m^2 - x (1 - x) \Delta^2$. Note the absence of one-loop
contributions to the $E$ and $C$ form factors from the nucleon operators. It
is a mere consequence of a cancellation of the one-loop contribution (c) by
self-energy insertions into the external lines. We have minimally absorbed
ultraviolet divergences generated in $M_{2,2}$ into higher order counterterms
that results into a redefinition of their coefficients, with which they enter
in the effective Lagrangian $r_i \to r_i + {\rm coeff}\, L_\varepsilon$, with
$L_\varepsilon \equiv \frac{1}{\varepsilon} - \gamma_E + \ln 4 \pi$. So that
the right-hand side of $M_{2,2}$ has extra analytic terms
\begin{eqnarray*}
\frac{m^2}{(4 \pi f_\pi)^2} r_1 + \frac{\Delta^2}{(4 \pi f_\pi)^2} r_2 \, .
\end{eqnarray*}
Note, that one can also completely absorb analytical contributions present in
(\ref{IsovectorFFs}) into counterterms. This simply sets the renormalization
scheme prescription which has to be used for all observables once $r_i$ are
fitted to experimental data in one.

Another comment concerns $E_{2,2}$. Pion operators generate $1/M$-suppressed
contributions to the structure of interest. Therefore, by power counting
we have to add ${\cal O} (1/M)$ bilinear nucleon operators constructed
from one large and one small components of the nucleon field as well as
analogous terms stemming from the chiral Lagrangian. Lorentz invariance
fixes unambiguously their coefficients and no new low energy constants
arise. However, the effect of these contributions in form factors is
$\Delta^2$-independent and analytic in the pion mass. Thus they will
not computed by us presently, although can be anticipated to generate an
addendum $- 5 a_2^N g_A^2 m^3 / \left( 32 \pi f_\pi^2 M \right)$ on the
right-hand side of $E_{2,2}$ in Eq.\ (\ref{IsovectorFFs}) required by
momentum sum rules discussed below. Also, there arises a Foldy-like term
$(b^N_j - a^N_j) \Delta^2/(4 M^2)$ in $E_{j,j}$, similarly to the form
factor case \cite{BerFeaHemMei98}, from contributions of the small
nucleon field components in the heavy-mass expansion of relativistic
nucleon operators.

Sum rules for the total momentum and spin of the nucleon (pion) impose
constraints on the coefficients $a^N_2$ and $b^N_2$ ($a^\pi_2$),
\begin{equation}
a^{N, q}_2 + a^{N, g}_2 = 1
\, , \qquad
b^{N, q}_2 + b^{N, g}_2 = 1
\, , \qquad
a^{\pi, q}_2 + a^{\pi, g}_2 = 1
\, .
\end{equation}
The latter two equations imply that the total gravitomagnetic moment of
the nucleon vanishes, i.e., $B^q_{2,2} (0) + B^g_{2,2} (0) = 0$
\cite{Ter99,BroHwaMaSch00}.

The leading $(j,j)$-structures of the higher $j$-moments ($j > 2$) do not
receive non-analytic contributions in the momentum transfer at next-to-leading
order in the chiral expansion due to the absence of relevant pion operators.
Thus one gets
\begin{equation}
E_{j,j} (\Delta^2) = a_j^N + \dots \, ,
\qquad
M_{j,j} (\Delta^2) = b_j^N
\left(
1 - 3 \frac{g_A^2 m^2}{(4 \pi f_\pi)^2}
\ln \frac{m^2}{\Lambda^2_\chi} + \dots
\right) \, ,
\end{equation}
where ellipsis stand for, at least, $m^2/(4 \pi f_\pi)^2$-suppressed
analytic contributions from one-loop diagrams and counterterms.

{\bf 5.} In the isovector sector, the use of Eq.\ (\ref{MagicRelation})
results into
\begin{equation}
\Sym_{\mu_1 \mu_2}
{\rm tr} \left\{
\tau^a
\partial_{\mu_1} {\mit\Sigma} \ \partial_{\mu_2} {\mit\Sigma}^\dagger
\right\} = 0 \, .
\end{equation}
Its generalizations for even $j = 2 k$ are straightforward and read
\begin{equation}
\Sym_{\mu_1 \mu_2 \dots \mu_j}
{\rm tr} \left\{
\tau^a
\partial_{\mu_1} \dots \partial_{\mu_k} {\mit\Sigma}
\
\partial_{\mu_{k + 1}} \dots \partial_{\mu_j} {\mit\Sigma}^\dagger
\right\} = 0 \, .
\end{equation}
Using this property it is easy to convince oneself that pion operators
for even $j = 2 k$ are reduced to the ones involving an odd number of
total derivatives, translated in the momentum space to odd powers of the
momentum transfer,
\begin{eqnarray*}
\Sym_{\mu_1 \dots \mu_j}
\left\{
\sum_{l = 0}^{k - 1} a^{\pi; (j)}_{2 l} \, ( - i \partial)^{2 l + 1} \,
{\rm tr} \tau^a
\left\{
( - i \partial)^{k - l} {\mit\Sigma} \ (i \partial)^{k - l + 1} {\mit\Sigma}^\dagger
+
( - i \partial)^{k - l} {\mit\Sigma}^\dagger \ ( i \partial)^{k - l + 1} {\mit\Sigma}
\right\}
\right\}_{\mu_1 \dots \mu_j} \, ,
\end{eqnarray*}
which, therefore, cannot contribute to the matrix elements in question due to
the time-reversal condition alluded to above.

For odd $j = 2 k + 1$, an elementary consideration along the same line as
above leads to the following structure of the non-singlet pion operators
\begin{eqnarray}
{\cal O}^{\pi, a}_{\mu_1 \mu_2 \dots \mu_j}
\!\!\!&=&\!\!\! \frac{f_\pi^2}{2} \Sym_{\mu_1 \dots \mu_j}
\\
&\times&\!\!\!\left\{
\sum_{l = 0}^{k} a^{\pi; (j)}_{2 l + 1} \, ( - i \partial)^{2 l} \,
{\rm tr} \tau^a
\left\{
( - i \partial)^{k - l} {\mit\Sigma} \ (i \partial)^{k - l + 1} {\mit\Sigma}^\dagger
+
( - i \partial)^{k - l} {\mit\Sigma}^\dagger \ (i \partial)^{k - l + 1} {\mit\Sigma}
\right\}
\right\}_{\mu_1 \dots \mu_j}
\, . \nonumber
\end{eqnarray}
The leading nucleon operators are given in Eq.\ (\ref{BaryonOperators})
with $A = a$.

Thus, for the isovector gravitational form factors there no contributions
from coupling to pions while the nucleon operators read
\begin{eqnarray}
{\cal O}^{N, a}_{\mu_1 \mu_2}
\!\!\!&=&\!\!\!
a^N_2 (2 M) \Sym_{\mu_1 \mu_2}
v_{\mu_1} v_{\mu_2}
\bar N \tau^a_{\xi +} N
+
b^N_2 (2 M)
(- i \partial_\nu)
\Sym_{\mu_1 \mu_2}
v_{\mu_1} \bar N \tau^a_{\xi +} \frac{[S_{\mu_2}, S_\nu]}{M} N
\nonumber\\
&+&\!\!\!
c^N_2 \frac{1}{2 M}
\Sym_{\mu_1 \mu_2}
(- i \partial_{\mu_1}) (- i \partial_{\mu_2})
\bar N \tau^a_{\xi +} N \, ,
\end{eqnarray}
with $\tau^a_{\xi +} \equiv \frac{1}{2} \left( \xi \tau^a \xi^\dagger
+ \xi^\dagger \tau^a \xi \right)$.

Due to the absence of the pion cloud contribution at this order of $\chi$PT,
i.e., diagrams (a) and (b), no non-analytic dependence on the momentum
transfer arises. However, the diagrams (c) and (d) develop chiral
logarithms in the pion mass of the form
\begin{eqnarray}
E_{2, 2} (\Delta^2)
\!\!\!&=&\!\!\!
a^N_2
\left\{
1
-
\frac{m^2}{(4 \pi f_\pi)^2}
\left(
(3 g_A^2 + 1) \ln \frac{m^2}{\Lambda^2_\chi}
- g_A^2 - 1
\right)
\right\}
\, ,\nonumber\\
M_{2, 2} (\Delta^2)
\!\!\!&=&\!\!\!
b^N_2
\left\{
1
-
\frac{m^2}{(4 \pi f_\pi)^2}
\left( (2 g_A^2 + 1) \ln \frac{m^2}{\Lambda^2_\chi} - 1
\right)
\right\}
\, , \nonumber\\
C_{2} (\Delta^2)
\!\!\!&=&\!\!\!
c^N_2
\left\{
1
-
\frac{m^2}{(4 \pi f_\pi)^2}
\left(
(3 g_A^2 + 1) \ln \frac{m^2}{\Lambda^2_\chi}
- g_A^2 - 1
\right)
\right\}
\, .
\label{IsoScalarFFs}
\end{eqnarray}
We imply that one adds counterterms to the right-hand side of these equations
linear in $m^2$ whose (unknown) coefficients absorb minimally the ultraviolet
divergences stemming from loops, ${\rm coeff}\, L_\varepsilon$, and the Foldy-like
term as discussed after Eq.\ (\ref{IsovectorFFs}). The leading structures of the
higher moments $M_{j,j}$ and $E_{j,j}$, apart from the change of an overall
normalization $b_2^N \to b_j^N$ and $a_2^N \to a_j^N$, have the same dependence
on the chiral logarithms as in Eq.\ (\ref{IsoScalarFFs}).

{\bf 6.} In this Letter we used the framework of the heavy baryon chiral
perturbation theory in one-loop approximation in order to predict the small
momentum transfer dependence of the moments of generalized parton distributions.
At next-to-leading order in the chiral expansion only the first two moments
develop a non-analytic $\Delta^2$ behavior: isovector electromagnetic
form factors and isosinglet gravitational form factors. Higher moments
can have at most a linear $\Delta^2$-dependence at this order of $\chi$PT.

Although, the momentum and spin sum rules suggest that the total quark and
gluon anomalous gravitomagnetic moment is zero \cite{Ter99,BroHwaMaSch00},
the gravitomagnetic radius of the nucleon does not vanish and reads
\begin{equation}
\langle {\sl r}^2 \rangle
\equiv
6 \left.
\frac{d M_{2,2} (\Delta^2)}{d \Delta^2}
\right|_{\Delta^2 = 0}
= - 3 a^\pi_2 \frac{g_A^2}{(4 \pi f_\pi)^2}
\left\{
\ln \frac{m^2}{\Lambda_\chi^2} + 1
\right\}
+ 6 \frac{r_2}{(4 \pi f_\pi)^2} \, .
\end{equation}

A pragmatic application of our present results consists of their use for
extrapolation of experimental measurements of the angular momentum sum
rule (\ref{SpinRule}) to zero momentum transfer \cite{BelMulKir01} and
as a fit formula for lattice simulations of the parton angular momentum
by means of the nucleon gravitational form factors \cite{MatDonLiuManMul00}.
Our considerations can be extended to other single-nucleon observables
such as parity-odd generalized parton distributions as well as off-forward
quark transversity and tensor gluon form factors.

\vspace{0.5cm}

We would like to thank J.-W. Chen and T.D. Cohen for useful discussions.
This work was supported by the US Department of Energy under contract
DE-FG02-93ER40762.

\end{document}